\documentclass[twocolumn,pra,aps,superscriptaddress,showpacs]{revtex4}
\usepackage{xspace,amsmath,amsfonts,amsthm,amssymb,amsbsy,graphics,color,graphicx,epsfig}
\usepackage{epstopdf}
\usepackage{url,hyperref}
\newcommand\foreign[1]{{\it #1\spacefactor=1000}}
\newcommand\eg{\foreign{e.g.}}

\newcommand{\sq}[1]{\left[ {#1} \right]}
\newcommand{\tr}[1]{{\textrm {Tr}}\sq{#1}}
\newcommand{\smallfrac}[2]{\mbox{$\frac{#1}{#2}$}}
\newcommand{\half}{\smallfrac{1}{2}}
\newcommand{\bra}[1]{\langle{#1}|}
\newcommand{\ket}[1]{|{#1}\rangle}
\newcommand{\ip}[2]{\langle{#1}|{#2}\rangle}
\newcommand{\op}[2]{\ket{#1}\bra{#2}}
\newcommand{\expt}[1]{\langle{#1}\rangle}
\newcommand{\dg}{^\dagger}
\newcommand{\D}[1]{{\cal D}\sq{#1}}

\newcommand{\Hc}[1]{{\cal H}\sq{#1}}
\newcommand{\Hcl}[1]{\tilde{{\cal H}}\sq{#1}}
\newcommand{\Lc}[1]{{\cal L}_c\sq{#1}}
\newcommand{\Lcun}[1]{{\tilde{\cal L}}_c\sq{#1}}
\newcommand{\Luc}[1]{{\cal L}\sq{#1}}

\newcommand{\beq}{\begin{equation}} 
\newcommand{\eeq}{\end{equation}}
\newcommand{\bqa}{\begin{eqnarray}} 
\newcommand{\eqa}{\end{eqnarray}}
\newcommand{\nn}{\nonumber} 

\newcommand{\erf}[1]{Eq.~(\ref{#1})}
\newcommand{\srf}[1]{Sec.~\ref{#1}}

\newcommand{\hei}{Heisenberg }

\begin{document}

\newtheorem{theo}{Theorem}
\newtheorem{lemma}{Lemma}

\newcommand{\red}{\color{red}}
\newcommand{\blue}{\color{blue}}

\newcommand{\blk}{\color{black}}

\title{Rapid purification  of quantum systems by measuring in a feedback-controlled unbiased basis}

\author{Joshua Combes}
\author{Howard M. Wiseman}

\affiliation{Centre for Quantum Computer Technology, Australia}
\affiliation{Centre for Quantum Dynamics, Griffith University, 
Brisbane, Queensland 4111, Australia}

\author{Kurt Jacobs}
\affiliation{Department of Physics, University of Massachusetts at Boston, 100 Morrissey Blvd, Boston, MA 02125, USA}

\author{Anthony J. O'Connor}
\affiliation{School of Biomolecular and Physical Sciences, Griffith University, Brisbane, Queensland 4111, Australia}

\begin{abstract}
Rapid-purification by feedback --- specifically, reducing the mean impurity faster than by measurement alone --- can be achieved by choosing the eigenbasis of the density matrix to be unbiased relative to the measurement basis. Here we further examine  the protocol introduced by Combes and Jacobs [Phys.~Rev.~Lett.~{\bf 96}, 010504 (2006)]  involving continuous  measurement of the observable $J_z$ for a $D$-dimensional  system. We rigorously re-derive the lower bound $(2/3)(D+1)$ on the achievable speed-up factor, and also an upper bound, namely $D^2/2$, for all feedback protocols that use measurements in unbiassed bases. Finally we extend our results to $n$ independent measurements on a register of $n$ qubits, and derive an upper bound on the achievable speed-up factor that scales linearly with $n$. 
\end{abstract}

 \date{\today}
\maketitle 

\section{Introduction}\label{intro}
Many quantum information processing (QIP) tasks require pure states as a resource \cite{Divin}. Preparation of pure states may be a bottleneck in certain physical implementations of some QIP protocols (\eg~solid state systems). In these systems measurement can not be treated as instantaneous \cite{OxtWarWis05,OxtWisSun06, OxtGamWis08, JorKor06}, and are often slow compared to the timescale for unitary operations \cite{Pet05}. In such systems, a projective measurement is approached only in the limit of long measurement times, by integrating the measured current \cite{GamBraWal07}. There are demonstrable advantages associated with using the dynamics of the measured current to perform read-out or state estimation \cite{GamBraWal07}. Technology is approaching the point where it can implement useful feedback protocols \cite{KubKocSam09, exp_control1, exp_control2, exp_control3, exp_control4, exp_control5, exp_control6, exp_control7}.  Thus it makes sense to consider combining  continuous measurement with  quantum feedback \cite{WisMil10} to control the  purification process.  
Our goal is to increase the rate at which these systems can be purified in order to alleviate the bottleneck identified above.   We note that, by the addition of a unitary at the end of the measurement process, rapid purification is identical with rapid state preparation \cite{Jac04, WisBou08,ComWisSco10}.

The Von-Neumann entropy, $S(\rho)= -\tr{\rho \log \rho}$, for a quantum state $\rho$ is a natural choice for characterizing a procedure that purifies and/or cools the state of a quantum system. Unfortunately, the von Neumann entropy is not an easy quantity with which to perform calculations. Because of this the linear entropy, given by $L(t)= 1-\tr{\rho(t)^2} $, is often used to make the calculations more tractable. In this paper we use the linear entropy, also known as the \textit{impurity}, to characterize the purification process. Since unitary operations on a quantum system leave the eigenvalues of the density matrix unchanged, in order to reduce the entropy of a system it must be allowed to interact with a bath. It is this interaction which enables, and is common to, all purification and cooling procedures.

Until recently the majority of work in the field of state-purification had been based on what we will term `open-loop' control methods, such as algorithmic cooling \cite{cool} or dissipation engineering \cite{diseng}. In this approach the bath remains unmeasured or the results of the measurements are forgotten. As a result the system evolves according to a deterministic master equation, such as
\begin{eqnarray}\label{ME}
  d\rho=dt\Luc{\ell} \rho \equiv  -i dt [H(t),\rho] + 
  2\gamma \, dt\, \D{\ell}\rho, 
  \end{eqnarray}
where $\D{A} \rho \equiv A\rho A\dg -\half (A\dg A \rho + \rho A\dg A)$ \cite{WisMil10}, $H(t)$ is the system Hamiltonian, $\ell$ is a lowering operator and $\gamma$ is a relaxation rate. Consequently the control signals or interactions are not conditioned on measurement results, and thus can be completely determined prior to the cooling process. Cooling can be thought of as a version of {\em purification} in which the final state is also the ground state for the system.

In measurement-based purification schemes the bath must be measured. In this case the system can be purified without control fields, provided the system observable coupled to the bath has no degenerate eigenvalues, by measurement alone. The change to an observer's state of  knowledge of an individual system after a weak measurement of the system observable $X$ is described by the stochastic master equation (SME)~\cite{CMreview2, CMreview1} 
\begin{eqnarray}\label{SME}
 d\rho= dt \Lc{X} \rho \equiv  
  2\gamma \, dt\, \D{X}\rho +\sqrt{2\gamma}\,dW\,\Hc{X}\rho,  
  \label{eq:drho1}
\end{eqnarray} 
where  $\Hc{A} \rho \equiv  A\rho +\rho A\dg - \tr{(A\dg+ A )\rho}\rho$ \cite{WisMil10}, and $dW$ is the increment of a Wiener noise process~\cite{Jacobs10}. It should be noted that we have moved to a frame that has enabled us to factor out the Hamiltonian evolution. The {\em measurement strength}, $\gamma$, determines the rate at which information is extracted, and thus the rate at which the system is projected onto a single eigenstate of $X$~\cite{VanStoMab0605, StovanHMab0408}. (This means for times $\tau \gg \gamma^{-1}$ we may say that we have performed a projective measurement of the observable $X$.) The measurement  result in a small time interval $[t,t+dt)$ is  
\begin{eqnarray}
dR = \sqrt{4\gamma}\langle X(t) \rangle dt + dW(t),
\end{eqnarray}
where $dW$ is the same Wiener noise process that appears in Eq.(\ref{eq:drho1}), and $\expt{X(t)} = \tr{X\rho(t)}$. We will denote the continuous measurement record obtained by the observer integrated up until time $t$ as $R(t) = \int_0^t dR(t')$. 

Conceptually a continuous measurement is a sequence of weak measurements, in the limit where the strength of the measurements tends to zero and the repetition rate tends to infinity~\cite{CavMil87, CMreview2, CMreview1,WisMil10}. Feedback, if thought about in the framework of a sequence of weak measurements, amounts to the ability to perform a unitary operation in-between each measurement \cite{CavMil87}. While unitary operations by themselves do not change the entropy of a system, the average {\em reduction} in the linear entropy caused by a measurement depends not only on the measurement but also on the prior state of the system. Feedback protocols that increase the rate of purification usually involve choosing the applied unitary after each weak measurement so as to increase the entropy reduction generated by the {\em next} weak measurement \cite{DohJacJun01,FucJac01,Jac03,Jac04,ComJac06,JorKor06,WisRal06,RalGriHil06, Jac07,GriHilRal07,HilRal07,JacLun07, ChiJac08, ComWisJac08, HilRal08, ShaJac08, WisBou08, BelNegMol09, LiJac09, ComWisSco10}.  

Here we consider rapid purification feedback protocols that apply unitary operations to continually keep the eigenbasis of the density matrix unbiased with respect to the basis of the observable being measured, $X$. (Two bases of a $D$-dimensional system are unbiased with respect to each other if the inner product of every vector of the first basis with every vector of the second basis is equal to $1/\sqrt{D}$.) We will refer to this class of feedback protocols  as ``unbiased-basis" (UBB) feedback protocols. Specifically, the goal of these protocols is to increase the rate at which the {\em average} linear entropy of the system, $\langle L(t) \rangle$, decreases as a function of time. The average here is taken over all possible trajectories (measurement records) for the evolution of the system under the measurement (and feedback). For a single qubit, the UBB protocol is optimal \cite{Jac03,WisBou08,BelNegMol09}. UBB protocols were first examined for systems of arbitrary dimension in~\cite{ComJac06}. The purpose of the present work is to clarify and extend the results in~\cite{ComJac06}, as well as to derive upper and lower bounds on the performance of UBB protocols when applied to registers of qubits. 

In Section~\ref{qubits} we review rapid purification for a single qubit. Section~\ref{L_Dfb_sec} discusses previous results on qudit feedback in an unbiased-basis. And Section~\ref{sec_ub_lb} provides a unified formalism to derive both the upper and lower bound on purification for any UBB feedback on a qudit. Finally, we derive upper and lower bounds for UBB protocols applied to a register of qubits in Section~\ref{regsec}, and conclude with a discussion of the results. 

\section{Rapid Purification for a Single Qubit}\label{qubits}

In this section we review the UBB protocol for a single qubit (originally presented in~\cite{Jac03}). For a single qubit the UBB protocol is optimal -- a non-rigorous proof of this is given in~\cite{Jac03}, and a rigorous proof is given in~\cite{WisBou08} and confirmed in Ref.~\cite{BelNegMol09}. It is not presently known whether UBB protocols are optimal for systems of higher dimension. 

To calculate the factor by which a feedback protocol speeds up the purification of a system we divide the time it takes the feedback protocol to achieve a given value of $\langle L(t)\rangle$ by the time it takes a measurement in the absence of feedback to achieve this value \footnote{Another natural optimization would be to minimize the average time it takes for a qubit to reach a fixed purity; for details see Refs.~\cite{WisRal06, ComWisJac08}.}. In both cases we start the system in the maximally mixed state. We will refer to a measurement acting without feedback as a \textit{bare} measurement~\cite{Jacobs05}. 

\subsection{Purification from Measurement Alone}\label{secqubitnfb}
We will perform our analysis keeping the dimension of the system arbitrary, for later convenience. Consider a quantum system of dimension $D$ initially in the maximally mixed state $\rho(0) = \textbf{I}/D$, where $\textbf{I}$ is the $D\times D$ identity matrix. It is possible to obtain a closed form expression for the linear entropy as a function of the measurement record by using the linear trajectory formulation of continuous measurements~\cite{Lintraj1,Lintraj2,WisMil10} (for a simple introduction see~\cite{CMreview2}). This involves solving the linear version of the SME, which produces an unnormalized density matrix. The linear version of the SME in \erf{SME} is 
 \begin{eqnarray}\label{SMEunorm}
 d\tilde{\rho}=dt\Lcun{X}\tilde{\rho} \equiv   
 2\gamma \, dt\, \D{X}\tilde{\rho} +\sqrt{2\gamma}\,dR\,\Hcl{X}\rho .  
 \end{eqnarray}
Here $\Hcl{A} \rho \equiv  A\rho +\rho A\dg$, and the tilde over $\rho$ denotes the lack of normalization at all but the initial time. Because the initial state is $\tilde{\rho} = \rho(0) = \textbf{I}/D$, the density matrix commutes with $X$ at all times, and this makes obtaining the solution simple. This solution is~\cite{CMreview2} 
\begin{eqnarray}\label{rhotildejz}
 \tilde{\rho}(R,t)&=&\exp(-4\gamma X^2t)\exp(2\sqrt{2\gamma}XR(t))\textbf{I}/D , 
\end{eqnarray}
where, as before, $R(t)$ is the integrated measurement record. For a qubit, for which $D=2$, we take the observable to be $X=J_z=\sigma_z / 2$. The solution becomes 
\begin{eqnarray}\label{qubitrhotilde}
 \tilde{\rho}(R,t)=\frac{e^{-\gamma t}}{2}\left( \begin{array}{c c }
e^{\sqrt{2\gamma}R}      & 0        \\
0      & e^{-\sqrt{2\gamma}R}       \\
\end{array} \right). 
\end{eqnarray}
(Here we have dropped the time-dependence of $R$ for compactness.) 

The final normalized density matrix $\rho(R,t)$ is given by dividing $\tilde{\rho}(R,t)$ by its norm, $\mathcal{N} = \mbox{Tr}[\tilde{\rho}(R,t)]$. The probability density that we obtain the state $\rho(R,t)$ at time $t$, is given by $\mathcal{P}(R,t)= \mathcal{N}\exp{(-R^{2}/2t)}/\sqrt{2\pi t}$. The average impurity of the final state is thus given by averaging the impurity $L[\rho(R,t)]$ over the probability density $\mathcal{P}(R,t)$. For a single qubit this gives 
\begin{equation}
   \label{nfb_qubit}
   \expt{L(t)} = \frac{e^{-\gamma t}}{\sqrt{8\pi t}} \int_{-\infty}^{+\infty} \frac{e^{-R^2/2t}}{\cosh(\sqrt{2\gamma}R)} dR .
\end{equation}
While this integral has no analytic solution (to our knowledge), we can obtain the behavior in the long-time limit by noting that the integral contains two multiplied distributions. The distribution in the numerator is broad compared to the distribution in the numerator for $t\gg \gamma^{-1}$. Thus in this long time (LT) limit, the integral can be approximated by  $\int_{-\infty}^{\phantom{..}\infty} dR/\cosh(\sqrt{2\gamma}R)  =\pi/\sqrt{2\gamma}$, and we have 
\begin{equation}
  \label{qubitlongtime}
  \expt{ L(t)}_{\mathrm{LT}} = \frac{\pi e^{-\gamma t}}{\sqrt{16\pi\gamma t }}.
\end{equation}
The key result is that the impurity for a bare continuous measurement scales  asymptotically as $e^{-\gamma t}$. 

We note that recently Jordan and Korotkov \cite{JorKor06} generalized \erf{nfb_qubit} for an arbitrary initial state, $\rho_0=\rho(0)= \smallfrac{1}{2}(\mathbf{I}+x \sigma_x+y\sigma_y +z\sigma_z)$. Using linear trajectory theory as above one can show that the impurity decays as
\begin{equation}\label{nfb_qubit_jk}
   \expt{L(t)} = \frac{e^{-\gamma t} L(\rho_0)}{\sqrt{2\pi t}} \!\! \int_{-\infty}^{+\infty} \!\! \frac{e^{-R^2/2t}dR}{\cosh(\sqrt{2\gamma}R)+z_0\sinh(\sqrt{2\gamma}R)},
\end{equation}
where $z_0= \tr{\sigma_z \rho(0)}$. The asymptotic expression is 
\begin{equation}
  \expt{ L(t)} _{\mathrm{LT}}= \frac{\pi e^{-\gamma t} }{\sqrt{16\gamma \pi t}}\frac{2L(\rho_0)}{\sqrt{1-z_0^2}}.
\end{equation}

\subsection{Purification Using Feedback}\label{qubitfb}

For the initial state $\rho(0) = \mathbf{I}/2$, the measurement dynamics is symmetric with respect to rotations about the $z$-axis. Because of this we lose nothing by restricting our feedback protocol to rotations about the $y$ axis at some rate $\alpha(t)$. The SME for this situation is 
\begin{eqnarray} 
d\rho & = &  -i dt[ \alpha (t)J_y,\rho] +dt\Lc{J_z} \rho(t) . 
  \end{eqnarray} 
To simplify the calculations we assume that $\alpha(t)$ can be arbitrarily large compared to the measurement strength $\gamma$. This means that we can consider the action of the control in each infinitesimal time-interval, $[t,t+dt)$, as a unitary that generates a rotation through any desired angle.
This unitary is therefore of the form 
\begin{equation}
   U_t \equiv U(t+dt,t) = \exp{\left\{ -i \alpha(t) J_y  dt\right\} }. 
\end{equation} 
In this case the state following an infinitesimal time-step consisting of measurement and feedback is
\begin{eqnarray}
 \rho_{\mathrm{fb}}(t+dt) &=&U_t \{ \rho(t) +d \rho(t) \}U_t\dg\\
\nonumber&=&U_t\{ \rho(t) +dt\Lc{J_z} \rho(t)\}U_t\dg . 
\end{eqnarray} 
Note that up to a unitary transformation, which has no effect on the purity, this is equivalent to having the feedback change the measurement basis: 
\begin{eqnarray}
   \rho_{\mathrm{fb}}(t+dt) &=& \rho(t) +dt\Lc{\check{X}(t)}\rho(t),
\end{eqnarray} 
where $\check{X}(t+dt)=U_t\check{X}(t)U_t\dg$ and $\check{X}(0)= J_z$. This is a Heisenberg picture with respect to the control unitary. In what follows we will always assume, for the sake of simplicity, that the feedback changes the measurement basis, rather than the state of the system. 
  
To derive an expression for $\expt{L(t)}$, we begin by examining the first-order change in the linear entropy. This is 
\begin{eqnarray}
   dL & = & d\left[1-\tr{\rho^2}\right] = -\tr{d(\rho^2)} \nonumber \\ 
   & = & -\tr{2\rho d\rho + (d\rho)^2 } . 
\end{eqnarray}
Here we must keep the second-order term in $d\rho$, because $\rho(t)$ is stochastic, and $(dW)^2 = dt$~\cite{Jacobs10}. From \erf{SME}, with $X$ replaced by $\check{X}$, the change in impurity is thus 
\begin{eqnarray}\label{dl}
dL & = & - 8\gamma dt \{ \tr{\rho \check{X}\rho\check{X}}-2\tr{\check{X}\rho}\tr{\check{X}\rho^2}     \nonumber  \\
\nonumber  &&\phantom{- 2\gamma dt \{}+\tr{\rho \check{X}}^2\tr{\rho^2}\}\\
 &&-4\sqrt{2\gamma}dW\{\tr{\check{X}\rho^2}-\tr{\rho \check{X}}\tr{\rho^2}\}.
\end{eqnarray}
To obtain the greatest decrease in impurity in each time interval $[t,t+dt)$, we must now optimize over all unitaries $U_t$ to obtain the locally optimal $\check{X}(t)$. It was shown in~\cite{Jac03} that for a single qubit this is achieved by choosing the eigenbasis of $\check{X}$ to be {\em unbiased} with respect to the eigenbasis of $\rho$. Each infinitesimal measurement disturbs this unbiased relationship, and thus feedback is required to maintain it. Since the disturbance to the basis of $\rho$ is proportional to $dW$ (rather than $dt$), to keep the bases {\em perfectly} unbiased requires that $\alpha(t)$ be arbitrarily large.  

If the bases of $\check X(t)$ and 
$\rho(t)$ are unbiased, then the bases of $\check X(t)$ and $\rho(t+dt)$ will only be infinitesimally biased, so that the necessary feedback unitary $U(t,t+dt)$ will be infinitesimally different from $\textbf{I}$. This is required for physically reasonable feedback. Mathematically, however, it is simpler to imagine the case where $\rho(t)$ is diagonal in the $J_z$ basis. In this case, $\check X(t)$ will be obtainable from 
$J_z$ by a finite unitary rotation.

We choose $\check{X} = J_x$, so that the unitary that transforms $X$ to $\check{X}$ via $\check{X}=TJ_zT\dg$, is 
\begin{equation}
T \equiv \exp{\left( i \frac{\pi}{2}J_{y} \right)}=  \frac{1}{\sqrt{2}}\left( \begin{array}{c c }
\phantom{-}1      & \phantom{-}1        \\
-1      & \phantom{-}1       \\
\end{array} \phantom{-}\!\!\!\right).
\end{equation} 
One can show that if $\check{X}$ is traceless, and unbiased with respect to $\rho$, then regardless of the dimension $\tr{\check{X}\rho}=0$ and $\tr{\check{X}\rho^2}=0$ (the derivation is given in Appendix~\ref{Xii0}). This considerably simplifies the expression for $dL$, which is now 
\begin{eqnarray}\label{gen_dpurity}
  dL & = &  -8\gamma\tr{\check{X}\rho \check{X}\rho}dt.
\end{eqnarray}
For a single qubit it turns out that $\tr{\check{X}\rho \check{X}\rho} = L/4$ and we obtain the very simple equation 
\begin{eqnarray}
  dL & = & -2\gamma dt L.
\end{eqnarray}
Evolution of the impurity is thus deterministic, and is given by 
\begin{eqnarray}
 L(t)&=&e^{- 2\gamma t} L(0). 
   \label{qubitfbL}
\end{eqnarray}
Note that the evolution of $L$ is only perfectly deterministic under the assumption that the observable $\check{X}$ is perfectly unbiased with respect to the density matrix.
In the long-time limit we obtain the speedup factor from Eqs. (\ref{qubitlongtime}) and (\ref{qubitfbL}). Denoting the time taken by a bare measurement to reach a given value of $L$ as $t_{\textrm{bare}}$, and that for the feedback protocol as $t_{\textrm{fb}}$, we equate $L(t_{\rm fb})=L(t_{\rm bare})$ and solve for the ratio $t_{\rm fb}/t_{\rm bare}$. Doing so gives
\begin{equation}\label{kurt_ratio}
  \frac{t_{\textrm{fb}}}{t_{\textrm{bare}}}=\frac{1}{S}=\frac{1}{2}+\frac{\ln{\sqrt{16\pi\gamma t_{\textrm{bare}}}}}{2\gamma t_{\textrm{bare}}}-\frac{\ln{2\pi}}{2\gamma t_{\textrm{bare}}}.
\end{equation}
For sufficiently large $t_{\textrm{bare}}$ (equivalently, a sufficiently small target impurity) the second and third terms are insignificant, and we obtain 
\begin{equation}\label{kurt_speedup}
  S =  \frac{t_{\textrm{bare}}}{t_{\textrm{fb}}} = 2.
\end{equation}
Numerical calculations show that for shorter times (higher target impurities) the speedup factor is always less than this. Thus the largest possible speedup for a single qubit is a factor of $2$. 

\section{qudit purification using feedback}\label{L_Dfb_sec}
\subsection{Purification from Measurement Alone}
\label{L_nfb_sec} 

To analyze UBB feedback protocols for $D$-dimensional systems, we will need the asymptotic evolution of the impurity for a bare measurement of $J_z$ for arbitrary $D$. In~\cite{ComJac06} it was shown only that at long times $L\sim e^{-\gamma t}$. A more detailed analysis, to be presented elsewhere~\cite{ComWisUN09}, shows that the asymptotic behavior is 
\begin{equation}\label{nfb_2_full}
   \expt{L_{2}(t)}_{\mathrm{LT}} = \frac{2(D-1)}{D}\frac{\pi e^{-\gamma t}}{\sqrt{16\gamma t \pi}}.
\end{equation}
Interestingly this is merely the expression for a single qubit, \erf{nfb_qubit}, multiplied by a factor that depends on the dimension of the system.
\subsection{The UBB protocol}\label{sec_ubbfb}
In this section we examine the generalization of the qubit rapid purification algorithm to $D$-dimesional systems (qudits), which was first proposed in \cite{ComJac06}. Recall that an UBB protocol for a $D$ dimensional system involves using feedback to continually adjust the measured observable $\check{X}(t)$ so that its eigenbasis remains unbiased with respect to the density matrix. 
Note that for these protocols we require only a single basis that is unbiased with respect to $\rho$ --- we do not require a complete set of mutually unbiased bases. 

The change in the impurity in a single infinitesimal time-step is once again given by Eq.(\ref{gen_dpurity}): 
\begin{eqnarray}\label{dgen_dpurity}
 dL&=&- 8\gamma\tr{\check{X}\rho \check{X}\rho}dt . 
\end{eqnarray}
However, in this case the right-hand side is no longer a simple function of $L$. Further, the evolution of $L$ need no longer be deterministic; $L$ is now coupled to other functions of $\rho$, and these will still evolve stochastically. It will be useful in what follows to write the right-hand side explicitly in terms of the matrix elements of $\check{X}$ in the elementary basis (the eigenbasis of $\rho$). Denoting this basis as $\{ |i\rangle \}$, $i = 1,\ldots,N$, we have 
\begin{eqnarray}
   \label{gen_dpurity2}
 dL & = &  -8\gamma dt \sum_{i,j} |\check{X}_{i,j}|^2\lambda_i\lambda_j . 
\end{eqnarray}
where $\lambda_j$ is the eigenvalue of $\rho$ associated with the eigenstate $|j\rangle$.

In~\cite{ComJac06} a lower bound was derived on the performance of UBB protocols, by considering a protocol in which simultaneous measurements are made of all observables that can be obtained from $\check{X}$ by permuting the basis vectors. This was possible because the evolution of $L$ under this protocol can be solved. Under the $D!$ simultaneous measurements, the total change in $L$, $dL_{\textrm{tot}}$, is simply the sum of the $dL$'s due to each measurement: 
\begin{eqnarray}
   \label{gen_dpurity_perms}
 dL_{\textrm{tot}} & = &  -\frac{8\gamma dt}{D!}\sum_{m=1}^{D!}\tr{{ \check{X}_m \dg \rho \check{X}_m \dg\rho}}   
\end{eqnarray}
where the $\check{X}_m{= P_{m}\dg \check{X} P_{m} }$ are $D!$ permutations of the operator $\check{X}$. 
Since at least one of the observables $\check{X}_m$ in the sum in Eq.(\ref{gen_dpurity_perms}) must give a $dL$ that is at least as large as the average over all the $\check{X}_m$, the performance of this protocol is a lower bound on the performance of protocols that employ a single optimized observable $\check{X}(t)$. An intricate calculation \cite{ComJac06} shows that \erf{gen_dpurity_perms} can then be rewritten in a remarkably simple way:
\begin{eqnarray}  
   \label{gen_dpurity_perms2}
   dL_{\rm tot}& = & -8\gamma dt \frac{(D-2)!}{D!}\tr{X^2} L.
\end{eqnarray}
 In Appendix~\ref{jzsumproof} we give a new, and much more detailed, proof of the process to obtain the relation in Eq.(\ref{gen_dpurity_perms2}) from \erf{gen_dpurity_perms}, and in Sec.~\ref{regsec} we will use this to obtain a lower bound on UBB protocols for a register of qubits. When $X=J_z$, $\tr{X^2}= \tr{J_z^2}=D(D^2-1)/12$, and \erf{gen_dpurity_perms2} becomes
\begin{eqnarray}
dL_{\rm LB}&=&-\smallfrac{2}{3}\gamma t(D+1)L\label{dimp_jz_feedback}\label{dL_perms_lb}.
\end{eqnarray}
The subscript LB indicates that this increment is a lower bound on $|dL|$ for any unbiased basis feedback, that is $|dL_{\rm LB}|\le |dL[\rho]|$. 

\section{Upper and lower bounds for rapid purification using UBB feedback}\label{sec_ub_lb}
We now introduce a new method to determine the lower bound obtained in~\cite{ComJac06}, \erf{dL_perms_lb} here, and a similar method that allows us to obtain an upper bound. 

The lower bound in Ref.\cite{ComJac06} was obtained by averaging over all permutations $P_m$ of the measured observable. We now note that this procedure renders $\expt{dL}$ invariant to such permutations. In light of this, and by analogy with the technique employed in Ref.~\cite{ComWisJac08}, we introduce a density matrix for which $dL$ is invariant under the permutations $P_m$: 
\begin{equation}
   \label{evrhof}
\rho_{\mbox{\scriptsize F}} = \mathrm{diag}\left(1-\Delta,\smallfrac{\Delta}{D-1},\smallfrac{\Delta}{D-1},\ldots,\smallfrac{\Delta}{D-1}\right).
\end{equation}
We call this the ``flat state", as it has one large eigenvalue and the remaining eigenvalues are equal in magnitude (``flat"). For any $D$-dimensional state $\rho$, with an impurity $L[\rho]$, we can always find a $\Delta$ so that  $L[\rho_{\mbox{\scriptsize F}}] = L[\rho]$. 

Intuitively the state in which $dL$ is {\em most} sensitive to permutations for a fixed value of $L$ is
\begin{equation} 
  \label{evrho2}
  \rho_2 = \mathrm{diag}(1-\Delta',\Delta',0,\ldots,0).
\end{equation}
We will refer to this state as the ``binary" distribution. Once again, for a given $\rho$ we can always find a $\Delta'$ so that  $L[\rho_2] = L[\rho]$. 

We now derive upper and lower bounds on UBB protocols by showing that when $L[\rho] = L[\rho_{\mbox{\scriptsize F}}] = L[\rho_2]$, and when each of these three density matrices have a permutation applied that maximizes $dL$, 
\begin{equation}\label{boundonL}
|\expt{dL[\rho_{\mbox{\scriptsize F}}]}| \leq |\expt{dL[\rho]}| \leq |\expt{dL[\rho_2]}| . 
\end{equation}

\subsection{The Lower Bound via the flat distribution}
\label{SEC_lb_rhof}

In \srf{L_Dfb_sec} we showed that the lower bound on the change in impurity is $dL_{\rm LB}=-\smallfrac{2}{3}\gamma t(D+1)L$. That is, $|dL_{\rm LB}| \leq |dL(\rho)|$. Now we show that $dL_{\rm LB}=dL(\rho_{F})$. We first note that the impurity of $\rho_{\mbox{\scriptsize F}}$ can be written as 
\begin{equation}
   \label{impflat}
    L[\rho_{\mbox{\scriptsize F}}]  = (D-1) \left[ \frac{2\Delta{(1-\Delta)}}{D-1} + (D-2)\frac{\Delta^2}{(D-1)^2} \right] .
\end{equation}
Equation (\ref{impflat}) will enable us to factor out the impurity in the following working. Substituting $\rho_F$ into \erf{gen_dpurity2} to calculate $dL(\rho_{F})$ (or $dL_{F}$ in shorthand notation) and then writing $dL_{F}$ in the form of \erf{impflat} we have
\begin{eqnarray}\label{lb}
\nn  dL_F&=&-8\gamma dt \left ( \frac{2\Delta(1-\Delta)}{(D-1)}\sum_{r\neq p}|\check{X}_{rp}|^2\right . \\
  &&\phantom{-8\gamma dt m}\left.+\frac{\Delta^2}{(D-1)^2}\sum_{r\neq p,c\neq p}|\check{X}_{rc}|^2\right ),
\end{eqnarray}
where the largest eigenvalue $1-\Delta$ is associated with some particular eigenstate $\ket{p}$. It is easy to see that $ dL_{F}$ is invariant under the transformation $dL[\rho_{F}]=dL(P\dg_{m}\rho_{F}P_{m})$ for all the $D!$ permutations $P_{m}$ labeled by $m$. The permutation invariance of $\rho_{F}$ already implies that $dL(\rho_{F})=(1/D!)\sum_{m=1}^{D!}dL(P_{m}\dg \rho_{F}P_{m})=dL_{\rm tot}=dL_{\rm LB}$. Nevertheless we persue the simplification of \erf{lb} by noting that for any unbiased-basis 
\begin{eqnarray}\label{ubb_sum}
\sum_{r\neq p}|\check{X}_{rp}|^2&=&(D+1)(D-1)/12,
\end{eqnarray}
as shown in Appendix \ref{ubbproof}. From this it follows that
\begin{eqnarray}
\sum_{r\neq p,c\neq p}|\check{X}_{r,c}|^2= \frac{D+1}{12}(D-1)(D-2).
\end{eqnarray} 
Thus it is possible to write \erf{lb} so that it has the same form as \erf{impflat}, simplifying the resulting expression gives:
\begin{equation}\label{impflat_fb}
 dL_F= -8\gamma dt \frac{(D+1)}{12} L[\rho_{F}]= -\smallfrac{2}{3}(D+1)\gamma dt L(t).
\end{equation}

It should be noted that technically we have not independently rederived the lower bound in \erf{dimp_jz_feedback}; rather we have shown that for all unbiased bases and for all possible permutations $P_{m}$ of $\rho_{F}$'s basis and for all impurities $dL_{\rm LB}=dL_{F}$. 

\subsection{The Upper Bound via the binary distribution}\label{SEC_ub_rho2}
 We now show that $ |dL[\rho]| \leq |dL[\rho_2]|$. Substituting $\rho_{2}$ in to \erf{gen_dpurity2} gives 
 $dL_{2}= -8\gamma dt[2(1-\Delta')\Delta' |\check{X}_{r,c}|^2]$. This expression {\em is} sensitive to the arrangement of the eigenvalues of $\rho_{2}$; accordingly $dL[\rho_{2}]\neq dL(P\dg_{m}\rho_{2}P_{m})$ for most permutations. 
 The impurity for $\rho_{2}$ can be written as
 \begin{equation}
 L[\rho_{2}]= 2(1-\Delta)'\Delta'= L[\rho]=\sum_{r\neq c,c\neq r}\lambda_{r}\lambda_{c}.
 \end{equation}

Using these  relations we find for the optimal permutation that $ dL(\rho_2)= -8\gamma dt[2(1-\Delta')\Delta'\max_{mn}|\check{X}_{mn}|^2]=-8\gamma dt \max_{mn}|\check{X}_{mn}|^2L[\rho]$. To prove $ |dL(\rho)| \leq | dL(\rho_2)|$ we need to prove that 
\begin{eqnarray}\label{ub_ineq}
\sum _{r,c}\lambda_{r}\lambda_{c}|\check{X}_{rc}|^2 \le \sum _{r,c}\lambda_{r}\lambda_{c}\max_{mn}|\check{X}_{mn}|^2.
\end{eqnarray}
This is trivially true since all the $\lambda_{i}$'s are positive. 
All that remains is to bound the $\max_{mn}|\check{X}_{mn}|^2$ in any unbiased-basis. 

A general unitary that transforms the basis $\ket{k}$ to an unbiased-basis is $T\ket{n}=\sum_{-j}^j\frac{1}{\sqrt{D}}\exp{(i\phi^{(n)}_k)}\ket{k}$. It is possible to rewrite  $|\check{X}_{mn}|^2$ as  $|\check{X}_{mn}|^2=|\expt{ m|T\dg J_zT|n}|^2$ so that
\begin{eqnarray}
 \nn \max_{mn}|\check{X}_{mn}|^2&\le &\max_{\{\phi^{(n)}_k\},\{\theta^{(m)}_k\}}\frac{1}{D^2}\left |\sum_{k=-j}^j e^{i(\phi^{(n)}_k-\theta^{(m)}_k)}k\right |^2 \\ &\le& \frac{1}{D^2}\left |\sum_{k=-j}^j k\right |^2.\label{gen_dl_lb}
\end{eqnarray}
For even $D$ \erf{gen_dl_lb} evaluates to $D^2/16$; for odd $D$ it evaluates to $D^2/16 -1/8 +1/(16D^2)$. For large $D$ (say $D>5$), $D^2 /16$ is a good approximation for both even and odd $D$. To find the lower bound on the decrease in impurity it is important to remember that two matrix elements contribute to the sum: $\max_{m,n}|\check{X}_{mn}|^2\lambda_{m}\lambda_{n}$, and $\max_{m,n}|\check{X}_{nm}|^2\lambda_{n}\lambda_{m}$. For large $D$ the impurity under the two eigenvalue distribution is then
\begin{eqnarray}\label{eigenvalueising2gen}
 dL_{\mathrm 2}&\leq &-8\gamma dt \frac{D^2}{16}  2\lambda_0\lambda_1=-\gamma dt \frac{D^2}{2}   L_2(t).
\end{eqnarray}
The dependance for this matrix element is only on the dimension of the system. Thus the speed-up upper bound for any unbiased-basis feedback is
\begin{eqnarray}\label{S_ub_all_mub}
S_{\mathrm 2} \leq \frac{D^2}{2},
\end{eqnarray}
for $D\gg 1$ and $t \gg \gamma^{-1}$.

\section{A Register of qubits}\label{regsec}

We now generalize UBB feedback protocols to the case of a register of $n$ qubits, where each qubit is independently and continuously measured. Instead of one observable $X$, we now have $n$, given by $X^{(r)}= I^{(1)}\otimes I^{(2)}\otimes \ldots \sigma_z^{(r)}\ldots \otimes I^{(n)} $, where $r$ labels the $r$th qubit. The SME describing such a measurment is 
\begin{equation}\label{sme_reg}
  d\rho = \sum _r 2\kappa \, dt\,\D{X^{(r)}}\rho+\sqrt{2\kappa}\,dW^{(r)}\Hc{X^{(r)}}\rho.
\end{equation}
The combined state of the $n$ qubits exists in a $D=2^n$ dimensional Hilbert space.

\subsection{Purification from Measurement Alone}
In this section we will not analyse the no-feeback case to the same level of rigor as we did in section \ref{secqubitnfb}, but rather, we rely upon the intuition gained from that analysis. 
For simplicity, consider first a two-qubit register with uncorrelated qubits. The state of the system is $\rho=\rho_1\otimes\rho_2$. The impurity for such a state is
\begin{equation}
L^{(2)} = 1-\tr{(\rho_1\otimes\rho_2)^2}= 1-\tr{\rho_1^2}\tr{\rho_2^2},
\end{equation}
where the superscript ``$(2)$'' on $L$ signifies the number of qubits in the register. For very pure states it is natural to parameterize the eigenvalues of the $r$th qubit, $\rho_r$, as $\{\lambda_0^r = 1-\Delta_r,\lambda_1^r=\Delta_r\}$ where the convention that $\Delta_r\ll 1$ still holds.
To first order in $\Delta$ we have
\begin{equation}
 L^{(2)} = 1-(1-2\Delta_1)(1-2\Delta_2)
\sim 2(\Delta_2  +\Delta_1).
\end{equation}
Because the qubits are initially uncorrelated, and are independently monitored, it is reasonable to assume that $\Delta_1$ and $\Delta_2$ are of the same order, so $L^{(2)}\sim 2(2\Delta)$. When $\Delta\ll 1$ then $L=2 (1-\Delta)\Delta\approx 2\Delta$ so $L^{(2)}\approx 2L$. At long times, for $n$ qubits we expect that
\begin{equation}
\expt{L^{(n)}}_{\rm LT}\sim n\expt{L^{(1)}}_{\rm LT}.
\end{equation}
We can verify this using the analytic expression for qubit impurity, \erf{nfb_qubit}. 
Recalling that the purity may be written as $P = 1-L$, we have $\expt{P^{(n)}}= \expt{P^{(1)}}^n$, so  

\begin{equation}\label{nfbreg}
  \expt{ L^{(n)}(t)} = 1-\left(1- \frac{e^{-4\kappa t}}{\sqrt{8\pi t}} \int_{-\infty}^{\infty} \frac{e^{-R^2/(2t)}dR}{\cosh(\sqrt{8\kappa}R)}  \right)^n.
  \end{equation}
Using the long-time expression for a qubit, \erf{qubitlongtime}, we obtain a long-time analytical expression for $L$ in the absence of feedback for a register of $n$ qubits
\begin{equation}\label{nfbreg}
   \expt{ L^{(n)}(t)}_{\rm LT} \sim1-\left(1- \frac{n\pi e^{-4\kappa t}}{\sqrt{64\pi\gamma t }}\right)=  \frac{n\pi e^{-4\kappa t}}{8\sqrt{\pi\kappa t }}.
\end{equation}
To compare this expression to the qubit expression, \erf{qubitlongtime}, $n$ is set to $1$ and $\kappa=\smallfrac{1}{4} \gamma$.

\subsection{Lower bound for purification using UBB feedback on a register of qubits}\label{SEC_reg_lb}
The change in impurity for a register of qubits in an unbiased basis is
\begin{eqnarray}
 dL&=&- 8\kappa dt \sum_{r=1}^n \tr{ \check{X}^{(r)}\rho \check{X}^{(r)}\rho}\\
 &=&- 8\kappa dt \sum_{r=1}^n\sum_{i,j=0}^{(D-1)} |\check{X}_{i,j}^{(r)}|^2\lambda_i\lambda_j \label{dl_reg_xij}.
\end{eqnarray}

Using $\rho_F$ to calculate the lower bound for a UBB protocol in a register appears to be difficult. Instead we will derive the lower bound using the same method presented in Appendix \ref{jzsumproof}.   To do this we will again introduce the feedback in the \hei picture so that $\check{X}^{(r,m)}= P_m T X ^{(r)} T\dg P_m \dg$. As before the $T$'s are conditional unitaries that introduce the unbiasedness (between $\rho$ and $X^{(r)}$) and the $P_m$'s also retain their meaning as permutations. For a register being measured in a basis that is unbiased with respect to the logical basis, with a randomly changing permutation, the change in impurity is 
\begin{eqnarray}
  \nonumber dL&=&\sum_{r=1}^n\sum_{m=1}^{D!}- 8\kappa dt  \tr{\check{X}^{(r,m)}\rho \check{X}^{(r,m)}\rho}
\end{eqnarray}
where $D = 2^n-1$. After performing a similar procedure to the one found in Appendix \ref{jzsumproof} we find 
\begin{eqnarray}
dL&=& -8\kappa dt nD(D-2)!L(t).
\end{eqnarray}
Thus we find the impurity of the state undergoing feedback of the above form decreases as 
\begin{eqnarray}\label{reg_ub}
L^{(n)}(t)&=&  e^{-8\kappa nt/(D-1)}L(0).
\end{eqnarray}
The asymptotic speed-up factor is
\begin{eqnarray}\label{regspeed}
S= \frac{2n}{D-1}.
\end{eqnarray}

For $n=1$ (i.e. a qubit) the speed-up factor $S$ is $S =2$, which agrees with the result from Ref.~\cite{Jac03}. When $n=2$, the speed-up is $S = 4/3$. This is comparable to the speed-up found for the locally optimal rapid measurement (RM) protocol in \cite{ComWisJac08} where the predicted speed-up in the long-time limit was $S_{\rm RM}\approx1.4$. (The feedback in the rapid measurement protocol permutes the eigenvalues of $\rho$ in the logical basis to decrease a different measure of impurity. Thus at all times the state eigenbasis and the measurement basis commute.) Unfortunately when $n=3$, $S =6/7$, which is a slow-down. This slow-down trend continues for all $n\ge 3$, and for large $n$ the slow-down is $\sim n2^{-n+1}$.

One can interpret \erf{regspeed} as a lower bound on an UBB algorithm for a register of qubits in much the same way as we did in Section \ref{sec_ubbfb}. Equation (\ref{regspeed}) may also be interpreted as an all-permutation deterministic purification protocol, although it is only useful in a two qubit register.

\subsection{Upper bound for purification using UBB feedback on a register of qubits}

As before it is possible to rewrite  $|\check{X}_{mn}^{(r)}|$ as  $|\check{X}_{mn}^{(r)}|=|\expt{ m|T\dg X^{(r)}T|n}|$ so that
\begin{eqnarray}
\max_{mn}|\check{X}_{mn}^{(r)}|&\le &\max_{\{\varphi_k\}}\frac{1}{D}\left |\sum_{k=0}^{D-1} e^{i\varphi_k}(-1)^{f(k,r)}\right |,
\end{eqnarray}
where $f(k,r)$ is a function that appropriately determines the sign of the diagonal elements of $X^{(r)}$. 
Thus
\begin{eqnarray}\label{Xcheck_2eval}
\max_{mn}|\check{X}_{mn}|&\le&
1,
\end{eqnarray}
and 
\begin{equation}\label{eigenvalueising2gen_reg}
\nonumber dL^{(n)}_{\mathrm 2}\le-8\kappa dt \sum_{r=1}^n L(t)=-8\kappa dt n L(t).
\end{equation}
At long times the speed-up upper bound for any UBB feedback in a register is 
\begin{eqnarray}\label{reg_ub}
S_{\mathrm 2} \le2n .
\end{eqnarray}
By substituting in $n=1$ we regain the result of Ref.~\cite{Jac03}. At present it is unclear if the bound in \erf{reg_ub} is tight. One reason why we expect this bound not to be tight comes from the upper bound on the rapid measurement protocol of Ref.~\cite{ComWisJac08}. The upper bound was shown to be $n$ and numerical results indicated it was $S_{\rm RM} =0.718 n$, which is significantly less than the upper bound.

\section{Discussion}

For a qudit we have bounded the speed-up in purification for any unbiased basis feedback by $\smallfrac{2}{3}(D+1)\le S\le D^2/2$. Prior to this work only the lower bound was known for the qudit. In Ref.~\cite{ComWisSco10} it was shown that by applying a unitary, chosen at random from the unitary group $\mathfrak{U}(D)$, to a qudit during the measurement process (random feedback) one could attain $S=\smallfrac{2}{3} D$ asymptotically. This suggests UBB feedback is unnecessary for a speed-up $O(D)$, although it remains an open question whether the actual speed-up for UBB protocols is $O(D)$ or $O(D^{2})$.

We also examined UBB feedback applied to a register of $n$ qubits. We have shown that the speed-up in purification is bounded by $2n/(2^{n}-1)\le S\le 2n$. Prior to this work there were no results for purification of a register of qubits using UBB feedback. 

Unlike the rapid measurement (RM) protocols of \cite{ComWisJac08} the UBB protocols presented in this paper are not locally optimized in time for decreasing $dL$. Additionally it is not clear if UBB protocols allow one to obtain information about the preparation procedure, unlike RM protocols which may be used for state estimation.

In the future we plan to investigate locally optimal protocols for the impurity (and other measures of mixedness), and examine rapid purification using a particular unbiased basis. Also, very recently Ruskov {\em et al.}. \cite{RusKorMol10} have considered monitoring a single qubit in three unbiased bases simultaneously. The relation between their work and ours remains to be explored.

{\em Acknowledgements:} The authors thank Andy Chia and Andrew Scott for helpful discussions. JC and HMW are supported by the Australian Research Council (FF0458313 and CE0348250). KJ is supported by the National Science Foundation, under Project No. PHY-0902906. 
\appendix
\section{Proof that $\check{X}_{ii}=0$ }\label{Xii0}
For a traceless operator $X$, which is diagonal in the computational basis, we can find a transformation $T$ that will make it unbiased with respect to the computational basis. The transformation has the form  $\check{X} = TXT\dg$
\begin{eqnarray}
\nonumber\check{X}_{il}&=& \sum_{j,k=1}^{D}T_{ij}X_{jk}T_{kl}\dg.
\end{eqnarray}
Now we wish to examine the diagonal elements. Being mindful of the fact that  $T_{rs}\dg=T_{sr}^*$ we have
\begin{eqnarray}
\nonumber\check{X}_{ii} = \sum_{j,k=1}^{D}T_{ij}X_{jk}T_{ki}\dg&=& \sum_{j,k=1}^{D}T_{ij}X_{jk}T_{ik}^*= \sum_{j,k=1}^{D}T_{ij}T_{ik}^*X_{jk}.
\end{eqnarray}
Using the fact that $X$ is diagonal gives the expression
\begin{eqnarray}
\check{X}_{ii}&=&\sum_{j=1}^{D}T_{ij}T_{ij}^*X_{jj}.
\end{eqnarray}
Unbiasedness means that $T_{ij}T_{ij}^*=1/D$ so that
\begin{eqnarray}
\check{X}_{ii}&=& \frac{1}{D}\sum_{j=1}^{D}X_{jj}.
\end{eqnarray}
Recalling that $X$ is traceless, we thus have $\check{X}_{ii}=0$.

\section{Proof of \erf{gen_dpurity_perms}}\label{jzsumproof}

Denote eigenvalues and eigenvectors of $\rho$ in the basis which diagonalizes $\rho$ as $\rho \ket{i} = \lambda_i\ket{i}$.   Define $X$ as a Hermitian operator that is diagonal in this eigenbasis, and $\check{X} = TXT\dg$ as a transformed version of $X$ that is unbiased with respect to this eigenbasis. Now define $P_m$, for $m=1,\ldots, D!$, to be the $D!$ operators that give each of the possible permutations of the eigenbasis $\ket{i}$: for example, $P_m\ket{i}=\ket{m(i)}$ labels a particular permutation of the basis $\ket{i}$, while $P_m\dg\ket{i}=\ket{m^{-1}(i)}$ is an inverse permutation to $m$. Note that $P_m\dg P_m\ket{i}=P_m P_m\dg\ket{i}=\ket{i}$. Starting with Eqn. \ref{gen_dpurity_perms} one may write
\begin{eqnarray}
\nonumber S&=& \sum_{m=1}^{D!}\tr{P_m \check{X} P_m\dg \rho P_m \check{X} P_m\dg\rho} \\
 &=&\sum_{m=1}^{D!}\sum_{i=1}^{D}\bra{i}\check{X} P_m\dg \rho P_m \check{X} P_m\dg\rho P_m\ket{i}\label{step1}.
\end{eqnarray}

Consider the action of the permutation operators and the state on the basis $\ket{i}$
\begin{equation}
 P_m\dg\rho P_m\ket{i} =P_m\dg\rho\ket{m(i)}= \lambda_{m(i)}P_m\dg\ket{m(i)}= \lambda_{m(i)}\ket{i}.
\end{equation}
Using the notation $\check{X}\ket{i}=\ket{\check{X}i}$, we thus have
\begin{eqnarray}
 S&=&\sum_{m=1}^{D!}\sum_{i=1}^{D}\lambda_{m(i)}\bra{\check{X}i} P_m\dg \rho P_m \ket{\check{X}i} .
\end{eqnarray}
Now we use the completeness of the basis ($\ip{a}{b} = \sum_j \bra{a}(\op{j}{j})\ket{b}$) so that
\begin{eqnarray}
 S&=&\sum_{m=1}^{D!}\sum_{i,j=1}^{D}\lambda_{m(i)}\ip{\check{X}i}{j}\bra{j} P_m\dg \rho P_m \ket{\check{X}i} \\
 &=&\sum_{m=1}^{D!}\sum_{i,j=1}^{D}\lambda_{m(i)}\lambda_{m(j)}\ip{\check{X}i}{j}\ip{j}{\check{X}i} \\
&=&\sum_{m=1}^{D!}\sum_{i,j=1}^{D}\lambda_{m(i)}\lambda_{m(j)}|\ip{\check{X}i}{j}|^2.
\end{eqnarray}

Next we wish to sum over the $m$ label, which represents permutations of the basis. First we define
\begin{equation}
C_{ij}=|\ip{\check{X}i}{j}|^2 \sum_{m=1}^{D!}\lambda_{m(i)}\lambda_{m(j)},
\end{equation}
so that $S = \sum_{ij}C_{ij}$. Suppose $i=j$, then
\begin{equation}
C_{ii}=|\ip{\check{X}i}{i}|^2 \sum_{m=1}^{D!}\lambda_{m(i)}^2.
\end{equation}
There are $(D-1)!$ permutations that take $i\rightarrow p$ for a fixed $p \in [1,2...,D]$, so
\begin{equation}
  C_{ii}=|\ip{\check{X}i}{i}|^2(D-1)! \sum_{p=1}^{D}\lambda_{p}^2= |\ip{\check{X}i}{i}|^2(D-1)! \tr{\rho^2}.
\end{equation}

Now consider the case where $i\neq j$. Take any pair $q\neq p$ and look for permutations that take $i\rightarrow p$, $j\rightarrow q$. There are $(D-2)!$ of these permutations:
\begin{equation}
C_{ij}=|\ip{\check{X}i}{j}|^2(D-2)! \sum_{p\neq q}^{D}\lambda_{p}\lambda_{q}.
\end{equation}
Now examine $ \sum_{p\neq q}^{D}\lambda_{p}\lambda_{q}$ term:
\begin{eqnarray}
\nonumber(\sum_p^D\lambda_p)^2&=& (\sum_p^D\lambda_p)(\sum_q^D\lambda_q)\\
\nonumber &=& \sum_p^D\lambda_p^2+ \sum_{p\neq q}^{D}\lambda_{p}\lambda_{q}\\
(\tr{\rho})^2&=&\tr{\rho^2} + \sum_{p\neq q}^{D}\lambda_{p}\lambda_{q}.
\end{eqnarray}

As $\rho$ is normalized we find $ \sum_{p\neq q}^{D}\lambda_{p}\lambda_{q}=1-\tr{\rho^2}$. The simplified expression for \erf{step1}, so far, is
\begin{eqnarray}
\nonumber S&=&\sum_{i}C_{ii}+\sum_{i,j\neq i}C_{ij}\\ 
 \nn &=&\sum_i(D-1)! \tr{\rho}|\ip{\check{X}i}{i}|^2\\
  &&+\sum_{i,j\neq i} (D-2)!(1-\tr{\rho^2})|\ip{\check{X}i}{j}|^2.
\end{eqnarray}

Now we simplify the expression $\sum_{i\neq j} |\ip{\check{X}i}{j}|^2$. Using the Parseval relation for a vector $\Psi$, $\ip{\Psi}{\Psi} = \sum_{j=1}^D|\ip{\Psi}{j}|^2$, we find
\begin{eqnarray}
\nonumber \sum_{j=1}^D|\ip{\check{X}i}{j}|^2 &=&  \sum_{j\neq i}^D|\ip{\check{X}i}{j}|^2+\sum_{j=i}^D|\ip{\check{X}i}{i}|^2\\
\ip{\check{X}i}{\check{X}i} &=&  \sum_{j\neq i}^D|\ip{\check{X}i}{j}|^2+\ip{\check{X}i}{i}^2,
\end{eqnarray}
i.e. $ \sum_{j\neq i}^D|\ip{\check{X}i}{j}|^2 = \ip{\check{X}i}{\check{X}i}-\ip{\check{X}i}{i}^2$. The total expression is now
\begin{eqnarray}
\nn S&=& \sum_i^D(D-1)! \tr{\rho}|\ip{\check{X}i}{i}|^2\\&&\nn+ (D-2)!(1-\tr{\rho^2})\sum_{i=1}^D( \ip{\check{X}i}{\check{X}i}-\ip{\check{X}i}{i}^2).\\
\end{eqnarray}
Massaging the $\sum_{i=1}^D \ip{\check{X}i}{\check{X}i}$ term gives 
\begin{eqnarray}
\nonumber\sum_{i=1}^D \ip{\check{X}i}{\check{X}i}&=& \sum_{i=1}^D\bra{i}\check{X}\dg \check{X}\ket{i}=\tr{\check{X}^2}.
\end{eqnarray}
Recall that $\check{X} =TXT\dg$, thus $\tr{\check{X}\dg \check{X}} = \tr{(TXT\dg)\dg TXT\dg}= \tr{X^2}$. 

Now we examine the $\sum_{i=1}^D \ip{\check{X}i}{i}^2$ term. The eigenvectors and eigenvalues of $\check{X}$ are $\check{X}\ket{\Psi_b}=x_b\ket{\Psi_b}$ and $\ip{i}{\Psi_b}=1/\sqrt{D}$ by way of their unbiasedness. By inserting the identity we have
\begin{eqnarray}
\nonumber \ip{\check{X}i}{i}&=& \sum_b \ip{\check{X}i}{\Psi_b}\ip{\Psi_b}{i}\\
\nonumber &=& \smallfrac{1}{\sqrt{D}}\sum_b  \bra{i}\check{X}\ket{\Psi_b}\\
\nonumber &=& \smallfrac{1}{\sqrt{D}} \sum_b x_b\ip{i}{\Psi_b}\\
&=&\smallfrac{1}{D} \sum_b x_b
 =\tr{\check{X}}/D,
\end{eqnarray}
so $\sum_{i=1}^D\ip{\check{X}i}{i}^2=\tr{\check{X}}^2/D^2 $. The total expression is
\begin{eqnarray}
\nn S&=& (D-1)! \tr{\rho}\tr{\check{X}}^2/D^2\\ \nn&&+(D-2)!(1-\tr{\rho^2})(\tr{\check{X}^2}-\tr{\check{X}}^2/D^2).\\
\end{eqnarray}
Finally, recalling from Appendix \ref{Xii0}  that $\check{X}$ is traceless, we have
\begin{eqnarray}
 S&=& (D-2)!\tr{\check{X}^2}(1-\tr{\rho^2})
\end{eqnarray}
as required.

\section{Proof of \erf{ubb_sum}}\label{ubbproof}
Recall that $\sum_{r\neq0}^D|\check{X}_{r0}|^2= \sum_{r=0}^D|\check{X}_{r0}|^2$ as $|\check{X}_{ii}|=0$. So, 
\begin{eqnarray}
\nonumber \bra{k}\check{X}\dg \check{X}\ket{k}&=& \bra{k}(U J_z U\dg)\dg U J_z U\dg\ket{k}
= \bra{k}U J_z^2 U\dg\ket{k}.
\end{eqnarray}
Now insert the identity
\begin{eqnarray}
\nonumber \bra{k}\check{X}\dg \check{X}\ket{k}&=& \sum_{l,m}\bra{k}U\op{l}{l} J_z^2\op{m}{m} U\dg\ket{k}\\
&=& \sum_{l,m}\bra{k}U\op{l}{l} J_z^2\op{m}{m} U\dg\ket{k}.
\end{eqnarray}
The matrix $J_z^2$ is diagonal, so 
\begin{eqnarray}
\nonumber \bra{k}\check{X}\dg \check{X}\ket{k} &=& \sum_{l}\bra{l}J_z^2\ket{l}\bra{k}U\ket{l} \bra{l} U\dg\ket{k}\\
\nonumber  &=& \sum_{l}\bra{l}J_z^2\ket{l}|\bra{k}U\ket{l}|^2\\
 &=&\frac{1}{D}\sum_{l}\bra{l}J_z^2\ket{l}
=\frac{D^2-1}{12}.
\end{eqnarray}

\bibliographystyle{apsrev}

\end{document}